\documentclass[a4paper,11pt,twoside]{article}

\usepackage{geometry} 
\usepackage{amsmath}
\usepackage{amssymb}
\usepackage{color}
\usepackage{graphicx}

\usepackage{epstopdf}  

\def\eps{\varepsilon}

\newcommand{\TeV}{\,\mathrm{TeV}}
\newcommand{\GeV}{\,\mathrm{GeV}}

\newcommand{\ord}[1]{\mathcal{O}\left( #1 \right)}

\newcommand{\Fig}[1]{Fig.~\ref{fig:#1}}

\newcommand{\Eq}[1]{Eq.~(\ref{eq:#1})}
\newcommand{\eq}[1]{eq.~(\ref{eq:#1})}

\newcommand{\omi}[1]{}

\DeclareMathOperator{\im}{Im}
\DeclareMathOperator{\re}{Re}

\newlength{\myem}
\newcommand{\sep}[1]{#1}
\newcounter{mysubequation}[equation]
\renewcommand{\themysubequation}{\alph{mysubequation}}
\newcommand{\mytag}{\stepcounter{mysubequation}%
\tag{\theequation\protect\sep{\themysubequation}}}
\newcommand{\globallabel}[1]{\refstepcounter{equation}\label{#1}}

\makeatletter
\renewcommand{\section}{\@startsection{section}{1}{0em}%
        {-3.5ex \@plus -1ex \@minus -.2ex}%
        {2.3ex \@plus.2ex}%
        {\normalfont\large\bfseries}}
\renewcommand{\subsection}{\@startsection{subsection}{2}{0em}%
        {-3.25ex\@plus -1ex \@minus -.2ex}%
        {1.5ex \@plus .2ex}%
        {\normalfont\bfseries}}
\renewcommand{\subsubsection}%
        {\@startsection{subsubsection}{3}{0em}%
        {-3.25ex\@plus -1ex \@minus -.2ex}%
        {1.5ex \@plus .2ex}%
        {\normalfont\itshape}}
\makeatother

\newcommand{\MM}{\mathcal{M}^2}

\newcommand{\W}{\mathcal{W}}

\newcommand{\bsg}{B\to X_s\gamma}
\newcommand{\BRbsg}{\text{BR}(\bsg)}

\def\gsim{\lower.7ex\hbox{$\;\stackrel{\textstyle>}{\sim}\;$}}
\def\lsim{\lower.7ex\hbox{$\;\stackrel{\textstyle<}{\sim}\;$}}
\def\order#1{{\cal O}(#1)}

\def\beq{\begin{equation}}
\def\eeq{\end{equation}}
\def\bea{\begin{eqnarray}}
\def\eea{\end{eqnarray}}
\def\hc{{\rm h.c.}}

\newcommand{\SISSA}{SISSA/ISAS and INFN, I--34014 Trieste, Italy}
\newcommand{\CERN}{CERN, Theory Division,
  CH--1211 Geneva 23, Switzerland}

\newcommand{\preprintdate}{}
\newcommand{\preprintnumber}{CERN--PH--TH/2008-240\\ SISSA--77/2008/EP}
 
\newcommand{\titletext}{Hierarchical Soft Terms and Flavor Physics} 
\newcommand{\authortext}{\large Gian F. Giudice$^{\, a}$, Marco Nardecchia$^{\, b}$, Andrea Romanino$^{\, b}$
\medskip\\\em\normalsize 
$\mbox{}^a$ \CERN
\\[0.1\baselineskip] 
$\mbox{}^b$ \SISSA}
\newcommand{\abstracttext}{We study the framework of hierarchical soft terms, in which the first two generations of squarks and sleptons are heavier than the rest of the supersymmetric spectrum. This scheme gives distinctive predictions for the pattern of flavor violations, which we compare to the case of nearly degenerate squarks. Experiments in flavor physics have started to probe the most interesting parameter region, especially in $b\leftrightarrow s$ transitions, where hierarchical soft terms can predict a phase of $B_s$ mixing much larger than in the Standard Model.}

\title{
\normalsize
\begin{tabular}[t]{l}
\preprintdate\end{tabular}
\hspace*{\fill}
\begin{tabular}[t]{l}\preprintnumber\end{tabular}
\vspace{3\baselineskip}\\\Large\bfseries\titletext\bigskip}
\author{\begin{minipage}[t]{0.8\textwidth}
\normalsize\centering\authortext
\end{minipage}}
\date{}

\begin{document}

\bigskip
\maketitle
\begin{abstract}\normalsize\noindent
\abstracttext
\end{abstract}\normalsize\vspace{\baselineskip}



\section{Introduction}
\label{sec:intro}
The softly-broken supersymmetric Standard Model introduces new terms in the Lagrangian with non-trivial transformation properties under the flavor symmetry group. These terms appear in the squark mass matrices and the trilinear interactions
\bea
 {{\tilde Q}_{L}^\dagger}{{\cal M}_{Q_L}^2} {{\tilde Q}_{L}}+ 
 {{\tilde D}_{R}^\dagger}{{\cal M}_{D_R}^2} {{\tilde D}_{R}}+ 
 {{\tilde U}_{R}^\dagger}{{\cal M}_{U_R}^2} {{\tilde U}_{R}}+
\nonumber \\
\left( {\tilde D}_R^\dagger Y_D A_D  {\tilde Q}_{L} H_D +
{\tilde U}_R^\dagger Y_U A_U  {\tilde Q}_{L} H_U  + \hc \right) .
\label{eq:softt}
\eea
Here $Y_{D,U}$ are the Yukawa couplings and generation indices have been suppressed. We concentrate on the quark sector, but the extension to leptons is straightforward.

Fully generic flavor-breaking structures in the soft terms are ruled out by experimental constraints. However, these constraints can be used to identify the restricted class of allowed soft terms, providing useful guidelines for model building. A broad class of theories is singled out by the hypothesis of Minimal Flavor Violation (MFV)~\cite{mfv}, which states that any flavor violation originates from Yukawa couplings. The MFV hypothesis effectively suppresses new-physics contributions to most of the flavor-violating processes. However, in the search for new effects in $K$, $D$ and $B$ physics it is useful to consider departures from exact MFV.  Usually such departures are described in terms of a small expansion parameter that measures the breaking of the flavor group or of one of its subgroups.
Three especially interesting examples have been studied in the literature.

{\it 1) Degeneracy.} The starting point is the universality assumption~\cite{dg}, which states that ${\cal M}^2$ and $A$ in \eq{softt} behave as flavor singlets. A distortion from exact universality comes from additional contributions to ${\cal M}^2$ and $A$ which are fully generic in flavor space, but their size is characterized by a smaller mass scale, $\delta \tilde m$. The small expansion parameter is given by the ratio of these two scales, $\delta {\tilde m}^2/{\tilde m}^2$, \emph{i.e.}\ the ratio between the flavor-violating and flavor-symmetric terms. The rotation angles that diagonalize the squark mass matrices are generally large, because they are neither suppressed by the expansion parameter nor related to CKM angles. The suppression of flavor-violating amplitudes arises from the near degeneracy of the quark mass eigenstates.

{\it 2) Alignment.} The assumption is that quark and squark mass matrices are nearly simultaneously diagonalized by a supersymmetric field rotation, either in the down or in the up sector~\cite{nirs}. The bounds from the kaon system 
severely constrain the case in which ${\cal M}_{Q_L}^2$ is aligned along the up direction. The bounds on $D^0$--$\bar D^0$ mixing give important constraints on the alignment along the down direction~\cite{Nir:2007ac}. Correlations between quark and squark mass matrices leading to alignment are possible in models where some approximate flavor symmetry determines the form of Yukawa couplings and soft terms~\cite{nirs,modali}. Flavor alignment does not imply mass degeneracy of squarks. Thus, in this case the situation is exactly reversed with respect to the case of degeneracy. The suppression of flavor violating processes is due to the small squark mixing angles, while squark masses can be widely different.

{\it 3) Hierarchy.} The flavor structure of the first and second generation squarks is tightly constrained by $K$ physics. On the other hand, the upper bounds on the masses of the first two generations of squarks are much looser than for the other supersymmetric particles. Therefore one can relax the flavor constraints, without compromising naturalness, by taking the first two generations of squarks much heavier than the third~\cite{dins,dimg,Pomarol:1995xc,ckn}. As discussed in more detail in Section~\ref{sec:nat}, this procedure alleviates, but does not completely solve, the flavor problem and a further suppression mechanism for the first two generations must be present. However, it is not difficult to conceive the existence of such a mechanism which operates if, for instance, the soft terms respect an approximate U(2) symmetry acting on the first two generations~\cite{Pomarol:1995xc,u2}. In the case of hierarchy, the small expansion parameter describing the flavor violation is the mismatch between the third-generation quarks identified by the Yukawa coupling and the third-generation squarks identified by the light eigenstates of the soft-term mass matrix. This small mismatch can be related to the hierarchy of scales present in the squark mass matrix and to CKM angles. However, for the phenomenological implications we are interested in, we do not have to specify any such relation and we can work in an effective theory where the first two generations of squarks have been integrated out. Their only remnant in the effective theory is the small mismatch between third-generation quarks and squarks.

In this paper, we will revisit the properties of hierarchical soft terms, concentrating especially on their implications in flavor physics. We will show how the hypothesis of hierarchy predicts correlations between $\Delta F=1$ and $\Delta F=2$ processes which are different from the correlations found in scenarios with degeneracy. We will present the bounds on the expansion parameters of the hierarchy case and compare them with the case of degeneracy. As a particularly interesting example we will study the phase of $B_s$ mixing, for which there are some claims~\cite{dev,CKMfitter,hfag} that experiments have measured an excess with respect to the SM prediction. We will show that the case of hierarchy is compatible with much larger phases of $B_s$ mixing than the case of degeneracy, and thus a hierarchical squark spectrum has more room to explain the alleged effect.  

\section{Hierarchical Soft Terms and Naturalness}
\label{sec:nat}

The hypothesis of hierarchical soft terms states that the first two generations of squarks and sleptons are much heavier than the rest of the supersymmetric particles, assumed to lie near the electroweak scale. We will denote by ${\tilde m}_h$ the mass of the heavy squarks and sleptons and by ${\tilde m}_\ell$ the mass scale of the other ``light" supersymmetric particles. The original motivation of this hypothesis~\cite{ckn} is that  ${\tilde m}_h$ is more weakly bound by naturalness arguments than other supersymmetric parameters, because its radiative effect on the Higgs mass parameter $m_H^2$ is rather moderate. The leading effect comes from a one-loop renormalization of $m_H^2$ proportional to an induced hypercharge Fayet-Iliopoulos term
\beq
{\rm Tr}(Y{\tilde m}^2)= {\rm Tr}({\tilde m}_Q^2+{\tilde m}_D^2-2{\tilde m}_U^2-{\tilde m}_L^2+{\tilde m}_E^2).
\label{eq:fy}
\eeq
Assuming that soft terms are generated at the GUT scale, this term leads to a naturalness bound on ${\tilde m}_h$ just below the TeV scale~\cite{dimg}. Nevertheless, the term in \eq{fy} vanishes if, at some energy scale, scalar masses are universal or satisfy a GUT condition where hypercharge is embedded in a non-abelian group. Since the term in \eq{fy} is only multiplicatively renormalized, it will remain zero at any scale.

\begin{figure}
\begin{center}
\includegraphics[width=0.66\textwidth]{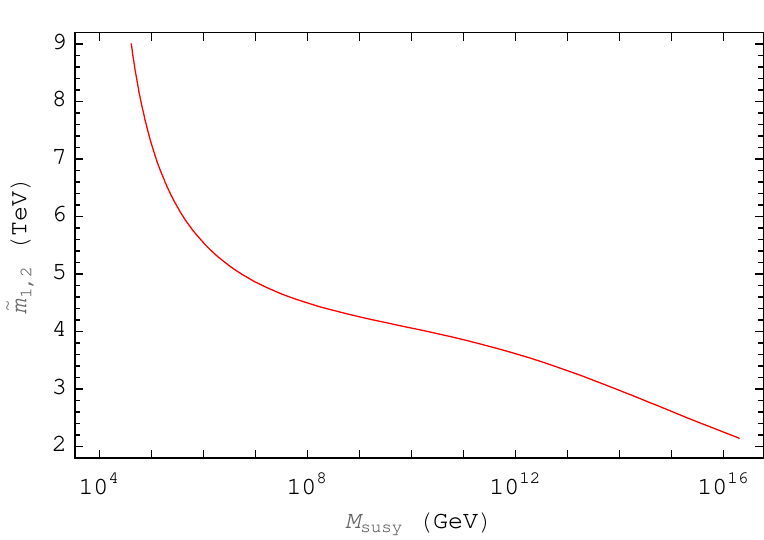}
\end{center}
\caption{Upper bound on ${\tilde m}_h$, assuming that first and second generation scalars are degenerate at a matching scale $M_{\rm susy}$. The bound corresponds to an upper limit $\Delta < 10$ on the fine-tuning parameter.}
\label{fig:nat}
\end{figure}

If the Fayet-Iliopoulos term vanishes, then the leading renormalization of $m_H^2$ proportional to ${\tilde m}_h^2$ comes from two-loop effects. In \Fig{nat} we show an upper bound on ${\tilde m}_h$, assuming that first and second generation scalars are degenerate at a matching scale $M_{\rm susy}$, where we start the renormalization group flow. The bound corresponds to an upper limit $\Delta < 10$ on the fine-tuning parameter $\Delta$~\cite{Barbieri:1987fn}, which is optimistic in the light of the present naturalness status of the supersymmetric SM. Still, multi-TeV squarks are allowed by naturalness. It is also possible to reach values of ${\tilde m}_h$ in the range of 10 TeV, but only if soft terms are generated at a very low scale $M_{\rm susy}$.

Another bound on the hierarchy of soft terms comes from the requirement that ${\tilde m}_h$ does not drive the squared masses of third-generation squarks to negative values, through its two-loop renormalization-group effect~\cite{2loopp}. This bound, although weaker than the previous one, is independent of naturalness arguments. Assuming complete degeneracy of the heavy states with mass ${\tilde m}_h$ and of the light states with mass ${\tilde m}_\ell$, the condition that color remains unbroken imposes ${\tilde m}_h/{\tilde m}_\ell \lsim 15$, if $M_{\rm susy}$ is close to the GUT scale. In the case of low $M_{\rm susy}$, where the effect is due to two-loop threshold effects not log enhanced, the bound becomes ${\tilde m}_h/{\tilde m}_\ell \lsim 25$. However, these bounds can be avoided by choosing appropriate boundary conditions of the soft terms at the scale $M_{\rm susy}$. For instance, all sfermions could be heavy at $M_{\rm susy}$, but Yukawa effects could dynamically bring the third generation to be light~\cite{fprun}. It is also possible to introduce new states that approximately cancel the two-loop renormalization-group contribution to ${\tilde m}^2_\ell$ proportional to ${\tilde m}^2_h$ and maintain the stability of the soft-term hierarchy against large radiative corrections~\cite{paz}.  

These upper bounds on ${\tilde m}_h$ have to be compared with the lower limits coming from flavor-violating effects in the $K$ system. Assuming that the heavy squark sector is neither degenerate nor aligned, we find the bound\footnote{These numbers are based on the analysis presented in Section~\ref{sec:bounds}. The effect of QCD corrections for heavy squarks has been considered in ref.~\cite{contino}.}
\beq
{\tilde m}_h > 35\TeV
\eeq
from the real part of the $\Delta S=2$ transition, and
\beq
{\tilde m}_h > 800\TeV
\eeq
from $\epsilon_K$.

This shows that the hypothesis of hierarchical soft terms is not sufficient to solve the flavor problem, unless one is willing to give up naturalness, in the spirit of Split Supersymmetry~\cite{split}, assuming that the first two generations of sfermions  are directly coupled to the supersymmetry-breaking sector. More concretely, we can retain naturalness and rely on a scheme for suppressing the flavor transitions in the heavy sector, as can be achieved by an approximate U(2) symmetry acting on the first two generations. 
In this respect, the hierarchical structure of soft terms can be a useful way of parametrizing supersymmetric theories which, for model-dependent reasons, have a certain separation of scales in the scalar sector. Moreover, hierarchical soft terms are interesting because they make specific predictions in flavor physics controlled by relatively few parameters related to physical quantities, like the mass hierarchy. As we will show, hierarchical soft terms offer a well-defined benchmark to be compared with the new experimental results in flavor physics.   

\section{Hierarchy versus Degeneracy in $\Delta F=1$ and $\Delta F=2$ Processes}
\label{sec:hiedeg}

Let us first consider the gluino contribution to a $\Delta F = 1$ process in the left-handed down quark sector, $d^L_i\rightarrow d^L_j$, neglecting for simplicity chirality changes. The amplitude of such a process is proportional to 
\begin{equation}
  \label{eq:ampl}
  A(\Delta F=1)\equiv f\left(\frac{\MM_D}{M_3^2}\right)_{d^L_id^L_j} =   \W_{d^L_i\tilde D_I}
  f\left(\frac{m^2_{\tilde D_I}}{M_3^2}\right) \W^*_{d^L_j \tilde D_I }.
\end{equation}
Here $f$ is a loop function, $M_3$ is the gluino mass and $\W$ is the unitary matrix diagonalizing the 6$\times$6 down squark squared mass matrix $\MM_D$ in a basis in which the down quark mass matrix is diagonal. We can simplify \eq{ampl} by using a perturbative expansion in the small off-diagonal entries of the squark mass matrix. It is often sufficient to keep the first order in the expansion. However, the second order can become important and even dominate in the case of 1--2 transitions, depending on the relative size of the 12 expansion parameter compared to the product of the 13 and 23 ones, and on the relative sizes of the sfermion masses. One important example of the case in which the second order dominates is the hierarchy case discussed below, in which the first order is suppressed because of the heaviness of the sfermions of the first two families. 
Then, \eq{ampl} becomes~\cite{BRS}
\begin{equation}
  \label{eq:PT}
  f\left(\frac{\MM_D}{M_3^2}\right)_{d^L_id^L_j} = \frac{\tilde m^2}{M_3^2}
  f\big(x_{d^L_i},x_{d^L_j}\big) \delta^{LL}_{ij} ,
\end{equation}
where $x_i \equiv m^2_i /M_3^2$,  $\delta^{LL}_{ij} \equiv \left(\MM_D\right)_{d^L_id^L_j}/\tilde m^2$, and
\begin{equation}
  \label{eq:rec}
  f(x,y) = \frac{f(x)-f(y)}{x-y} .
\end{equation}
The ``mass insertion'' $\delta^{LL}_{ij}$ is the expansion parameter and we have normalized it to a mass $\tilde m$ which can be chosen to be a typical scale of squark masses. This parameter effectively accounts (at first order) for the flavor transition.

The ``degenerate'' case is obtained in the limit in which the squark masses in the loop function coincide,
\begin{equation}
  \label{eq:D}
  m^2_{\tilde d^L_i} = m^2_{\tilde d^L_j}  \equiv \tilde m^2 .
\end{equation}
With this assumption, we obtain
\begin{equation}
  \label{eq:DMI}
  f\left(\frac{\MM_D}{M_3^2}\right)_{d^L_id^L_j} = x f^{(1)}(x)  \, \delta^{LL}_{ij}, 
  \qquad
\text{(degenerate case)}
\end{equation}
where $x = \tilde m^2/M_3^2$ and $f^{(n)}$ is the $n$-th derivative of the function. The $\delta$ parameters are in this case normalized to the universal scalar mass $\tilde m^2$. 

In the ``hierarchical'' limit, the contribution to the loop function in \eq{ampl} from the heavy squarks is negligible. Therefore \eq{ampl} becomes
\begin{equation}
  \label{eq:HMI}
  f\left(\frac{\MM_D}{M_3^2}\right)_{d^L_id^L_j}= 
  f(x) \, \hat\delta^{LL}_{ij}. \qquad
 \text{(hierarchical case)}
\end{equation}
Here $x = \tilde m^2/M^2_3$ as before, where now  $\tilde m^2$ is interpreted as the third-generation squark mass. We have defined $\hat\delta^{LL}_{ij} \equiv \W_{d^L_i\tilde b_L} \W^*_{d^L_j \tilde b_L}$. Note that $\hat\delta^{LL}_{a3} \approx - (\MM_D)_{d^L_ad^L_3}/\tilde m^2_{a}$, so that $\hat\delta^{LL}_{a3}$ is again a normalized mass insertion. Also, $\hat\delta^{LL}_{12} = \hat\delta^{LL}_{13}(\hat\delta^{LL}_{23})^*$. \Eq{HMI} can also be obtained from an extension of \eq{PT} to the second order in $\delta$.

\begin{figure}
\begin{center}
\includegraphics[width=0.66\textwidth]{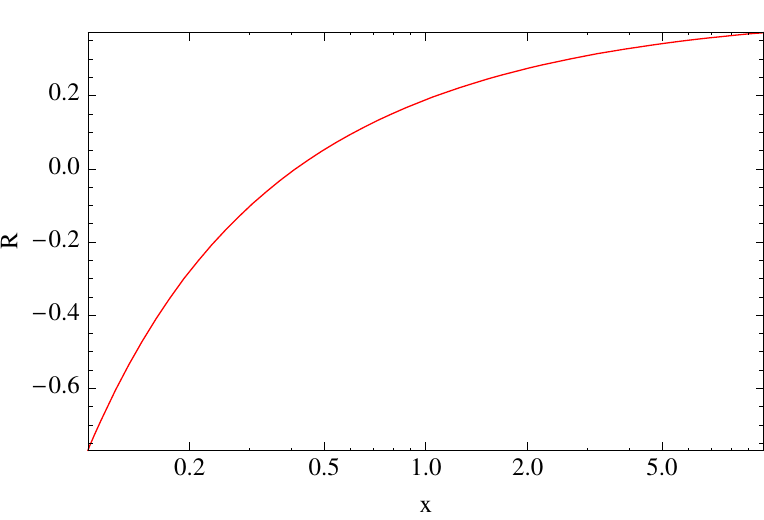}
\end{center}
\caption{Dependence of the factor $R = (g^{(3)}/6g^{(1)})(f/f^{(1)})^2$ on $x = \tilde m^2/M^2_3$. The loop functions enter the coefficients of the LL insertions in the $B_s$--$\bar B_s$ oscillation amplitude and in the $\bsg$ decay amplitude.}
\label{fig:ratio}
\end{figure}

Equations (\ref{eq:DMI}) and (\ref{eq:HMI}) show that for $\delta = \hat\delta$ the difference between the two schemes, the degenerate and the hierarchical one, is given by the order one difference between a function and its derivative. However, this $\ord{1}$ difference becomes larger when we consider $\Delta F = 2$ processes and turns out to affect the predicted correlation between $\Delta F = 1$ and $\Delta F = 2$. In fact, let us now consider the gluino contribution to a $\Delta F = 2$ $d^L_i\leftrightarrow d^L_j$ process. The amplitude is proportional to 
\begin{equation}
  \label{eq:MI2}
 A(\Delta F=2)\equiv \W_{d^L_i\tilde D_I} \W_{d^L_i\tilde D_J}
  g\left(\frac{m^2_{\tilde D_I}}{M_3^2}, \frac{m^2_{\tilde D_J}}{M_3^2}\right)
  \W^*_{d^L_j \tilde D_I} \W^*_{d^L_j \tilde D_J},
\eeq
where the loop function $g(x,y)$ is of the form\footnote{This decomposition follows from the form of the loop integral
$$
g(x,y)=\int dk \frac{G(k)}{(k^2-x)(k^2-y)}=\frac{1}{x-y} \int dk ~G(k)\left( \frac{1}{k^2-x}-\frac{1}{k^2-y}\right) \equiv \frac{g(x)-g(y)}{x-y}. \nonumber
$$}
\beq
g(x,y)=\frac{g(x)-g(y)}{x-y}.
\eeq
Expanding in the small off-diagonal elements of the squark mass matrix and assuming, as in the case of $\Delta F=1$, the dominance of $2\times 2$ transitions, we obtain that \eq{MI2} can be written as
\beq  
A(\Delta F=2)= \frac{\tilde m^4}{M_3^4}  {\hat g}\big( x_{\tilde d^L_i}, x_{\tilde d^L_j}\big) 
 (\delta^{LL}_{ij})^2,
\end{equation}
\beq
{\hat g}(x,y) =\frac{g(x,x)-2g(x,y)+g(y,y)}{(x-y)^2}.
\eeq
Thus, \eq{MI2} becomes
\begin{equation}
  \label{eq:DHMI}
  A(\Delta F=2) =
    \begin{cases}
    \displaystyle
    \frac{x^2}{3!} g^{(3)}(x) (\delta^{LL}_{ij})^2 &
    \text{(degenerate case)} \\[3mm]
    g^{(1)}(x) (\hat\delta^{LL}_{ij})^2 &
    \text{(hierarchical case).}
  \end{cases}      
\end{equation}
Therefore, if $\tilde m^2$ is the same in the two cases we find that the amplitudes for $\Delta F=1$ and $\Delta F=2$ processes satisfy the relation
\begin{equation}
  \label{eq:2vs1}
  \left.\frac{A(\Delta F = 2)}{[A(\Delta F = 1)]^2}\right|_{\text{degenerate}} =
  \frac{g^{(3)}}{6g^{(1)}} \left(\frac{f}{f^{(1)}}\right)^2
  \left.\frac{A(\Delta F = 2)}{[A(\Delta F = 1)]^2}\right|_{\text{hierarchical}}.
\end{equation}
This result is independent of the values of the mass insertions in the two cases. Partly due to the different factorials involved, the ratio $R = (g^{(3)}/6g^{(1)})(f/f^{(1)})^2$ is typically small, easily $\ord{10^{-1}}$ for $x=1$. As a consequence, the bounds on the $\Delta F=2$ processes inferred from $\Delta F=1$, or viceversa, may be significantly different in the two frameworks. The factor $R$ is shown in \Fig{ratio} as a function of $x = \tilde m^2/M^2_3$. The loop functions entering the factor $R$ plotted in \Fig{ratio} are the ones entering the coefficients of the LL insertions in the $B_s$--$\bar B_s$ oscillation amplitude and in the $\bsg$ decay amplitude. 

Another interesting point has to do with the relation between the $s\leftrightarrow d$, $b\leftrightarrow d$, $b\leftrightarrow s$ $\Delta F=2$ processes. In the degenerate case, such processes are proportional (for given chiralities and charge of the gaugino involved) to the a priori independent three quantities $\delta_{sd}^2$, $\delta_{bd}^2$, $\delta_{bs}^2$. A partial correlation among the three processes could in principle be generated by higher order contributions to the $s\leftrightarrow d$ transitions, {\it e.g.}
the ones proportional to $\delta_{sb}^2\delta_{bd}^2$. However, such contributions turn out to be always small. This is because of the limits on the two factors $\delta_{bs}$ and $\delta_{bd}$ and because the four-insertions $\delta_{sb}^2\delta_{bd}^2$ contribution is proportional to $(x^4/5!)\, g^{(5)}$, \emph{i.e.}\ it is suppressed by the factor 5! = 120. On the other hand, in the hierarchical case, a correlation does arise because $\hat\delta_{ds} = \hat\delta_{db}\hat\delta_{sb}^*/ |\W_{b\tilde b}|^2 \approx \hat\delta_{db}\hat\delta_{sb}^*$. Moreover, the higher-order contribution proportional to $\hat\delta_{db}^2\hat\delta_{sb}^{*2}$ is now proportional to $g^{(1)}(x)$, with no factorials involved. 

\section{The Flavor Structure for Hierarchical Soft Terms}
\label{sec:def}

In this Section we define the setting of hierarchical soft terms in greater detail. In order to obtain the expressions for the amplitude of a generic flavor process in the hierarchical case, it suffices to consider the case of a one-variable loop function, as in $\Delta F=1$ transitions. The generalization to more variables is straightforward. Let us then consider an amplitude whose dependence on sfermion masses comes through
\begin{equation}
  \label{eq:ampl2}
  f\left(\frac{\MM}{M^2}\right)_{Ai,Bj} =   \W_{Ai,I}
  f\left(\frac{\tilde m^2_I}{M^2}\right) \W^*_{Bj,I} . 
\end{equation}
In the expression above, $M$ is the mass of the relevant gaugino, $\MM$ is the $6\times 6$ sfermion squared mass matrix in the up squark, down squark, or charged slepton sector (the extension to sneutrinos is again straightforward), written in a basis in which the corresponding fermion mass matrix is diagonal and positive. The amplitude corresponds to a flavor transition between two fermions with chirality $A,B = L,R$ of families $i, j = 1,2,3$, $i\neq j$ and $\W$ is the unitary matrix diagonalizing $\MM$, so that $\W_{Ai,I}$ is the mixing between the fermion ``$Ai$'' and the $I$-th sfermion mass eigenstate, $I=1\ldots6$. 

According to our assumption, 4 out of the 6 squarks are much heavier than the others or the gaugino mass.
Their contribution to the loop function is then suppressed by the light-to-heavy ratio of squared masses $\tilde m^2_\ell/\tilde m^2_h$ (at least, up to logarithms) with respect to the contribution from third-generation squarks. However, for flavor transitions between quarks of the first two families, the exchange of third-generation squarks comes at the price of mixing angles, also suppressed by powers of the heavy mass scale $\tilde m^2_h$. Nevertheless, as shown in the Appendix, the contribution of heavy squarks to \eq{ampl2} is subdominant, as long as some GIM mechanism is operative in the first two generation squark sector. Since this must be the case in order to evade the strong constraints from $\epsilon_K$, we can neglect the effect of the heavy squarks in the summation of \eq{ampl2}. Alternatively, the assumption of neglecting the heavy-state exchange is justified when the first two generations of squarks are completely decoupled and the flavor mixing of the third-generation squarks are determined by quark rotation angles (see Appendix). 

We are then left with two light squarks with masses $\tilde m_{\ell_\alpha}$ and mixings $\W_{Ai,\alpha}$, where $\alpha=1,2$ is the index of the light eigenstates. This gives a total of 2+20 real parameters. However, since the mixings always appear in the combination $\W_{Ai,\alpha}\W^*_{Bj,\alpha}$, the overall phases of the mixing parameters (for any value of $\alpha$) do not affect \eq{ampl2} and the number of effective parameters is 2+18. 

This is still more general than needed. In fact, the decoupling of the first two sfermion families leads (under certain assumptions) to two additional constraints, as discussed in the Appendix. First, in the limit in which the 4 heavy sfermions decouple, the 2$\times$2 matrix $\W_{A3,\alpha}$ that diagonalizes the 2$\times$2 third-family sfermion mass matrix is approximately unitary. It is then always possible to describe it in terms of an angle $0\leq \theta\leq \pi/2$ and a phase $\phi$. The angle $\theta$ corresponds to the usual mixing angle between the two chiral components of third-generation squarks.  Second, the chirality-changing mixing is subdominant with respect to the chirality-conserving one, except within the third family. This means that the leading effect in any chirality-changing transition comes from the combination of a chirality-conserving one times a $\theta$-angle rotation. 

We are then left with 4 parameters describing the third generation squarks (${\tilde m}_{\ell_\alpha}$, $\theta$, $\phi$) and the four complex chirality-conserving ``insertions'' $\hat\delta^{LL}_{i3}$, $\hat\delta^{RR}_{i3}$, $i=1,2$ defined as follows:
\globallabel{eq:defins}
\begin{align}
\hat\delta^{LL}_{i3} &\equiv \sum_{\alpha=1,2} \W_{Li,\alpha}\W^*_{L3,\alpha} &
\hat\delta^{LL}_{3i} & = \hat\delta^{LL*}_{i3} \mytag \\
\hat\delta^{RR}_{i3} &\equiv \sum_{\alpha=1,2} \W_{Ri,\alpha}\W^*_{R3,\alpha} &
\hat\delta^{RR}_{3i} & = \hat\delta^{RR*}_{i3} . \mytag
\end{align}
Using the expression of the matrix $\W$ derived in the Appendix, at first order in the insertion $\hat \delta$,  
\eq{ampl2} becomes
\globallabel{eq:fLO2}
\begin{align}
f\left(\frac{\MM}{M^2}\right)_{Li,Lj} &= 
\left[\cos^2\theta ~ f\left(\frac{\tilde m^2_{\ell_1}}{M^2}\right) +
\sin^2\theta ~f\left(\frac{\tilde m^2_{\ell_2}}{M^2}\right) \right]
\hat\delta^{LL}_{i3} {\hat\delta^{LL*}_{j3}} 
\mytag \\
f\left(\frac{\MM}{M^2}\right)_{Li,Rj} &= 
\sin\theta\cos\theta e^{i\phi} \left[f\left(\frac{\tilde m^2_{\ell_1}}{M^2}\right) -
f\left(\frac{\tilde m^2_{\ell_2}}{M^2}\right) \right]
\hat\delta^{LL}_{i3} {\hat\delta^{RR*}_{j3}} \mytag \\ 
f\left(\frac{\MM}{M^2}\right)_{Li,L3} &= 
\left[\cos^2\theta ~f\left(\frac{\tilde m^2_{\ell_1}}{M^2}\right) +
\sin^2\theta ~f\left(\frac{\tilde m^2_{\ell_2}}{M^2}\right) \right]
\hat\delta^{LL}_{i3}
\mytag \\
f\left(\frac{\MM}{M^2}\right)_{Li,R3} &=
\sin\theta\cos\theta e^{i\phi} \left[f\left(\frac{\tilde m^2_{\ell_1}}{M^2}\right) -
f\left(\frac{\tilde m^2_{\ell_2}}{M^2}\right) \right] 
\hat\delta^{LL}_{i3} . \mytag
\end{align}

Equations~(\ref{eq:fLO2}) further simplify if the mixing angle $\theta$ is small, as in the case of the down squark sector in the moderate $\tan\beta$ regime. By taking, for simplicity, equal masses for the third generation squarks, $\tilde m_{\ell_1} \approx \tilde m_{\ell_2} \equiv \tilde m$, we obtain
\globallabel{eq:fLO4}
\begin{align}
f\left(\frac{\MM}{M^2}\right)_{Li,Lj} &=
f\left(x\right) 
\hat\delta^{LL}_{ij} 
\mytag \\
f\left(\frac{\MM}{M^2}\right)_{Li,Rj} &=
x\, f^{(1)}\left(x \right) 
\hat\delta^{LR}_{ij}  \mytag \\ 
f\left(\frac{\MM}{M^2}\right)_{Li,L3} &=
f\left(x \right) 
\hat\delta^{LL}_{i3}
\mytag \\
f\left(\frac{\MM}{M^2}\right)_{Li,R3} &=
x f^{(1)}\left(x \right) 
\hat\delta^{LR}_{i3} ,
\mytag
\end{align}
where $x = \tilde m^2/M^2$ and we have defined
\globallabel{eq:effins}
\begin{align}
\hat\delta^{LL}_{ij} &\equiv \hat\delta^{LL}_{i3} {\hat\delta^{LL*}_{j3}} \mytag \\[1.8mm]
\hat\delta^{LR}_{ij} &\equiv \frac{\MM_{L3,R3}}{\tilde m^2} 
\hat\delta^{LL}_{i3} {\hat\delta^{RR*}_{j3}} 
\qquad i,j=1,2 \mytag \\
\hat\delta^{LR}_{i3} &\equiv \frac{\MM_{L3,R3}}{\tilde m^2}
\hat\delta^{LL}_{i3} . \mytag
\end{align}
Here we have written $e^{i\phi}\sin\theta$ as $\MM_{L3,R3}/(\tilde m^2_{\ell_1} - \tilde m^2_{\ell_2})$. Equations~(\ref{eq:effins}) express two important results of the flavor structure of hierarchical soft terms. The flavor transition between the first two generations ($\hat\delta^{LL}_{ij}$) is determined by the product of the transitions involving the third generation ($\hat\delta^{LL}_{i3} {\hat\delta^{LL*}_{j3}}$). The chiral-violating flavor transitions ($\hat\delta^{LR}_{ij}$ and $\hat\delta^{LR}_{i3}$) are determined by the product of chiral-conserving flavor transitions and the chiral violation in the third family (${\MM_{L3,R3}}/{\tilde m^2}$).

\section{Bounds on Flavor-Violating Parameters}
\label{sec:bounds}

We now illustrate the bounds on the flavor-violating parameters $\hat \delta$ and $\delta$ in the hierarchical and degenerate cases, respectively. An early analysis of the hierarchical case was presented in ref.~\cite{Cohen:1996sq}. Our results for the LL insertions are summarized in Table~\ref{tab:bounds}. For definitess, here and below we set the $A$-terms to zero and we consider the case $\tilde m = M_3  = \mu$, with $\tilde m$ normalized to $350\GeV$. This choice allows a direct comparison with several results in the literature and is appropriate for the sbottom mass. For simplicity we use the same value for the stop mass, relevant in the case of $D^0$--$\bar D^0$ oscillations, although that is barely compatible with the Higgs mass bound. For sufficiently large $\tan\beta$, the leading chiral flip in the sbottom sector comes from $\mu v \tan\beta$. The limits on the RR insertions are the same, except the one from $\BRbsg$, which is much weaker. This is because the contribution of the LL insertion to the $\bsg$ amplitude interferes with the SM one, while the RR contribution does not. 

The bounds have been computed by constructing two-dimensional likelihood functions in the $\re{\delta}$--$\im{\delta}$ planes. Such functions have been obtained using a standard bayesian approach. The real and imaginary parts of the insertions are varied with flat distributions and the input parameters, summarized in Table~\ref{tab:inputs}, are varied according to their distributions. The likelihood function is then constructed from a fit of the relevant experimental values, also shown in Table~\ref{tab:inputs}. The expressions for the supersymmetry contributions to the Wilson coefficients in terms of the hierarchical insertions have been obtained from~\cite{BBMR,greub}. They have been used at the scale $\tilde m$ and then runned at lower scales according to~\cite{greub,magici,magici2}. 

\begin{table}
\begin{equation*}
\begin{array}{|c|}
\hline
\\
\textrm{$D_0-\bar{D}_0$ mixing}\\
\renewcommand{\arraystretch}{1.5}
\begin{array}{|c|c|}
\hline
\left|  \hat \delta^{LL}_{ut} \hat \delta^{LL*}_{ct}\right| < 
8.0 \times 10^{-3} \left( \frac{m_{\tilde{t}}}{350 \ \textrm{GeV}} \right) &
\left|  \delta^{LL}_{uc} \right| < 
3.4 \times 10^{-2} \left( \frac{m_{\tilde{q}}}{350 \ \textrm{GeV}} \right) \\
\hline
\end{array}\\
\\
B \to X_s \gamma \\
\renewcommand{\arraystretch}{1.5}
\begin{array}{|c|c|}
\hline
\big| \re \big( \hat \delta^{LL}_{sb} \big) \big| < 
2.2 \times 10^{-2} \left( \frac{m_{\tilde{b}}}{350 \ \textrm{GeV}} \right)^2 \left( \frac{10}{\tan \beta} \right) &
\left| \re \left( \delta^{LL}_{sb} \right)  \right| < 
3.8 \times 10^{-2} \left( \frac{m_{\tilde{q}}}{350 \ \textrm{GeV}} \right)^2  \left( \frac{10}{\tan \beta} \right) \\
\big| \im \big( \hat \delta^{LL}_{sb} \big)  \big|  < 
6.7 \times 10^{-2} \left( \frac{m_{\tilde{b}}}{350 \ \textrm{GeV}} \right)^2  \left( \frac{10}{\tan \beta} \right) &
\left| \im \left(\delta^{LL}_{sb} \right)  \right|  < 
1.1 \times 10^{-1} \left( \frac{m_{\tilde{q}}}{350 \ \textrm{GeV}} \right)^2  \left( \frac{10}{\tan \beta} \right) \\
\hline
\end{array} \\
\\
\Delta m_{B_s} \\
\renewcommand{\arraystretch}{1.5}
\begin{array}{|c|c|}
\hline
\big| \re \big( \hat \delta^{LL}_{sb} \big)  \big|  < 9.4 \times 10^{-2}  \left( \frac{m_{\tilde{b}}}{350 \ \textrm{GeV}} \right) &
\left| \re \left( \delta^{LL}_{sb} \right)   \right| <  4.0 \times 10^{-1} \left( \frac{m_{\tilde{q}}}{350 \ \textrm{GeV}} \right) \\
\big| \im \big( \hat \delta^{LL}_{sb} \big)  \big| < 7.2 \times 10^{-2} \left( \frac{m_{\tilde{b}}}{350 \ \textrm{GeV}} \right) &
\left| \im \left( \delta^{LL}_{sb} \right) \right|  <  3.1 \times 10^{-1} \left( \frac{m_{\tilde{q}}}{350 \ \textrm{GeV}} \right) \\
\hline
\end{array} \\
\\
B_d^0\text{--}\bar B_d^0 \text{ mixing} \\
\renewcommand{\arraystretch}{1.5}
\begin{array}{|c|c|}
\hline
\big| \re \big( \hat \delta^{LL}_{db} \big)  \big|  < 4.3 \times 10^{-3} \left( \frac{m_{\tilde{b}}}{350 \ \textrm{GeV}} \right)&
\left| \re \left( \delta^{LL}_{db} \right)  \right|  < 1.8 \times 10^{-2} \left( \frac{m_{\tilde{q}}}{350 \ \textrm{GeV}} \right) \\
\big| \im \big( \hat \delta^{LL}_{db} \big)  \big|  < 7.3 \times 10^{-3} \left( \frac{m_{\tilde{b}}}{350 \ \textrm{GeV}} \right) &
\left| \im \left(\delta^{LL}_{db} \right)   \right|  < 3.1 \times 10^{-2} \left( \frac{m_{\tilde{q}}}{350 \ \textrm{GeV}} \right) \\
\hline
\end{array} \\
\\
\Delta m_{K} \\
\renewcommand{\arraystretch}{2}
\begin{array}{|c|c|}
\hline
\sqrt{ \Big| \re \big( \hat{\delta}^{LL}_{db}  \hat{\delta}^{LL*}_{sb} \big)^2 \Big|} <  1.0 \times 10^{-2} \left( \frac{m_{\tilde{b}}}{350 \ \textrm{GeV}} \right) &
\sqrt{ \left| \re \left( \delta^{LL}_{ds} \right)^2 \right|} <   4.2 \times 10^{-2} \left( \frac{m_{\tilde{q}}}{350 \ \textrm{GeV}} \right)  \\
\hline
\end{array} \\
\\
\epsilon_{K} \\
\renewcommand{\arraystretch}{2}
\begin{array}{|c|c|}
\hline
\sqrt{ \Big| \im \big(  \hat{\delta}^{LL}_{db}  \hat{\delta}^{LL*}_{sb}  \big)^2 \Big|} < 4.4 \times 10^{-4} \left( \frac{m_{\tilde{b}}}{350 \ \textrm{GeV}} \right) &
\sqrt{ \left| \im \left( \delta^{LL}_{ds} \right)^2 \right|} < 1.8 \times 10^{-3}  \left( \frac{m_{\tilde{q}}}{350 \ \textrm{GeV}} \right)  \\
\hline
\end{array} \\
\\
\hline
\end{array}
\end{equation*}
\caption{Bounds on the LL insertions in the hierarchical and degenerate cases. The limits on the RR insertions are the same, except the one from $\BRbsg$, which is much weaker. The bounds are obtained at the 95\% CL from one-dimensional distributions defined as explained in the text.}
\label{tab:bounds}
\end{table}

The bounds on $s\leftrightarrow d$ transitions are obtained using the constraints from the kaon mass difference $\Delta m_K$ and the kaon mixing CP-violation parameter $\epsilon_K$. Because of the large theoretical uncertainty on the long-distance part of $\Delta m_K$, the absolute value of the supersymmetry contribution to $\Delta m_K$ has been allowed to be as large as its experimental value, with a flat probability distribution. For each parameter $\delta$ (degenerate or hierarchical, LL or RR) a combined two-dimensional likelihood function is first built in the $\re{\delta}$--$\im{\delta}$ plane. The likelihood for $\sqrt{|\re(\delta^2)|}$ (or $\sqrt{|\im(\delta^2)|}$) is then obtained as the section along the $\sqrt{|\im(\delta^2)|} = 0$ (or $\sqrt{|\re(\delta^2)|} = 0$) direction and is used to determine the 95\% CL limits shown in Table~\ref{tab:bounds}. The limit from $\Delta m_K$ is compatible with the limit in~\cite{GGMS}, whereas the limit from $\epsilon_K$ is stronger. This is because the allowed range for the supersymmetric contribution to $\epsilon_K$ is now smaller, in particular it is not anymore allowed to take values as large as the SM contribution. 

\begin{figure}
\begin{center}
\includegraphics[width=\textwidth]{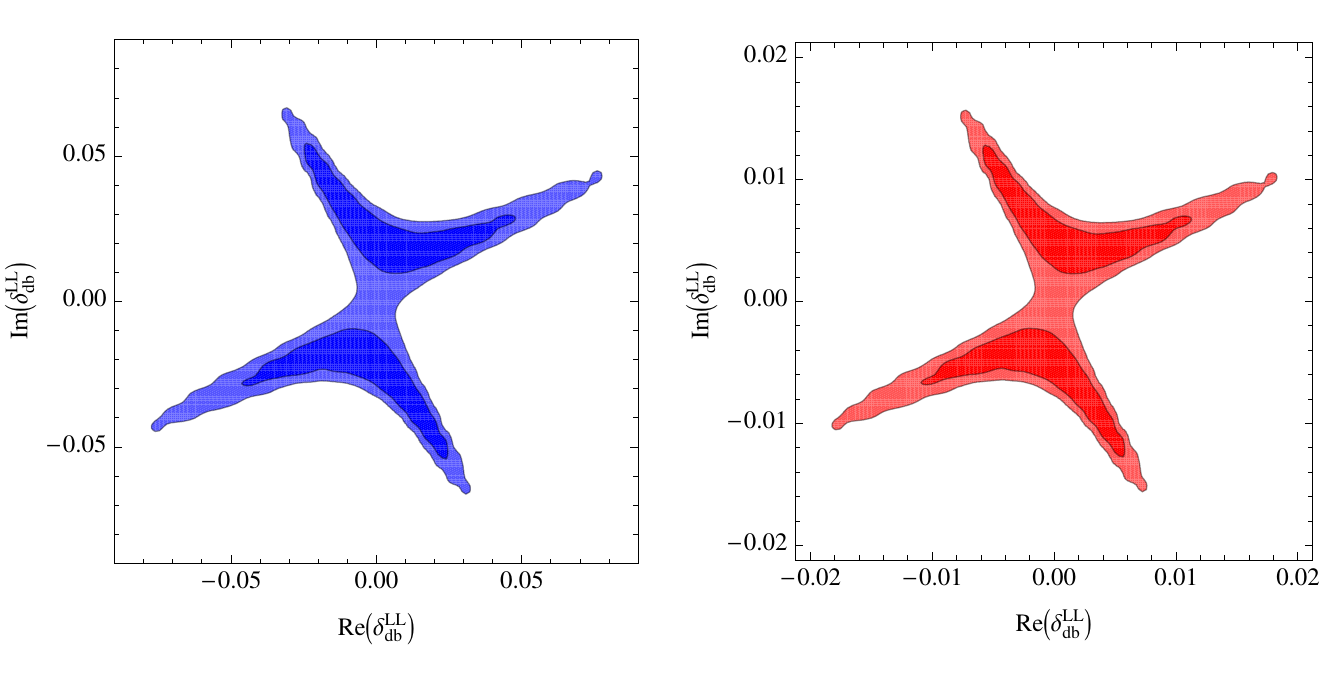}
\end{center}
\caption{95\% CL (light shading) and 68\% CL (dark shading) bounds on the real and imaginary parts of $\delta^{LL}_{db}$ (left, blue) and $\hat\delta^{LL}_{db}$ (right, red) from the measurements of $\Delta m_{B_d}$ for $\tilde m = M_3  = \mu = 350\GeV$.}
\label{fig:bd}
\end{figure}

The bounds on $b\leftrightarrow d$ transitions are obtained using the constraint from the $B^0_d$--$\bar B^0_d$ system mass difference $\Delta m_{B_d}$ and on the phase of the corresponding amplitude. Again, a two-dimensional likelihood is constructed. The corresponding 95\% CL and 68\% CL regions in the $\re{\delta}$--$\im{\delta}$ plane are shown in \Fig{bd}. The bounds on $\re{\delta}$ ($\im{\delta}$) in Table~\ref{tab:bounds} are obtained from the one-dimensional section of the two-dimensional likelihood corresponding to $\im{\delta} = 0$ ($\re{\delta} = 0$). Choosing $\im{\delta} = 0$ makes the limit on $\re\delta$ in Table~\ref{tab:bounds} much stronger than the size of the allowed region in the Figure. The corresponding constraint in Table~\ref{tab:bounds} should therefore be considered as optimistic. \Fig{bd} also shows that the point $\hat\delta^{LL}_{db} = 0$ ($\delta^{LL}_{db} = 0$) is excluded at more than $1\sigma$. This is a consequence of the mild deviation from the SM or MFV hypothesis observed in $b\leftrightarrow d$ transitions (see e.g.\ \cite{CKMfitter}).

\begin{figure}
\begin{center}
\includegraphics[width=\textwidth]{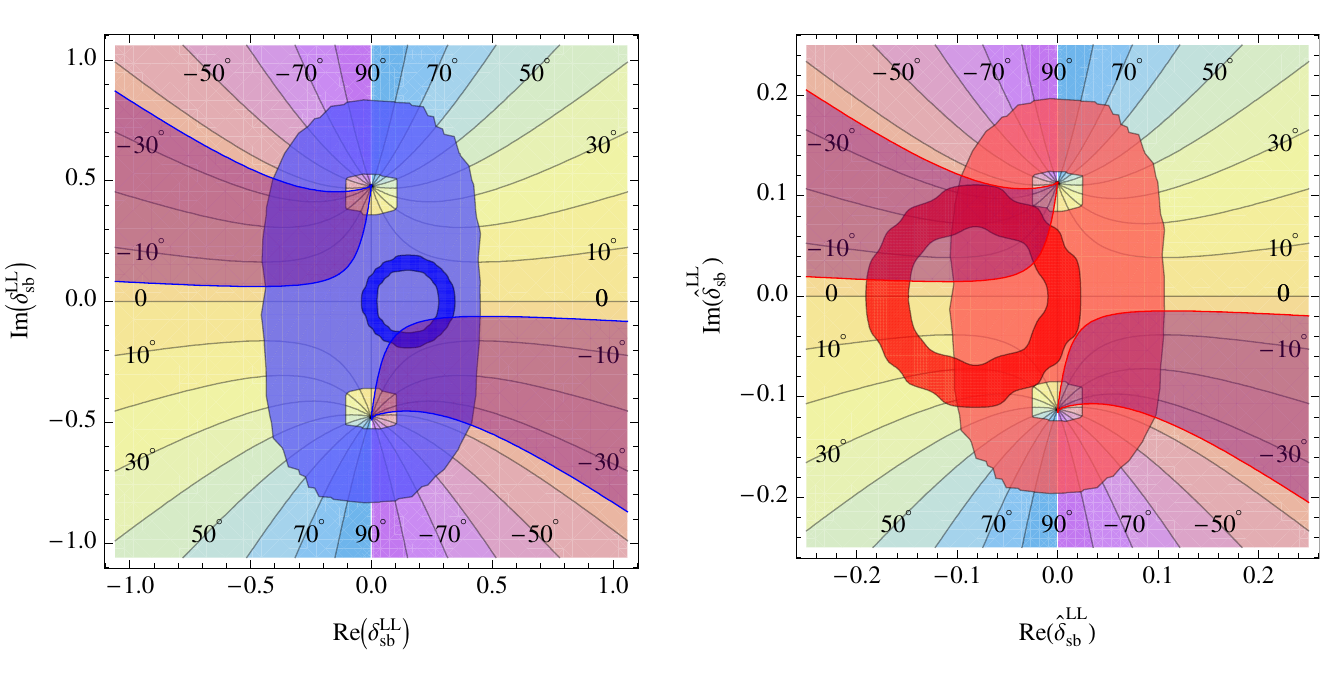}
\end{center}
\caption{95\% CL bounds on the real and imaginary parts of $\delta^{LL}_{sb}$ (left, blue) and $\hat\delta^{LL}_{sb}$ (right, red) from the measurements of $\Delta m_{B_s}$ (lighter shading) and $\BRbsg$ (darker shading) for $\tilde m = M_3  = \mu = 350\GeV$ and $\tan\beta =10$. Switching the sign of $\mu$ approximately corresponds to switching the sign of $\re (  \delta^{LL}_{sb})$ and $\re (  \hat \delta^{LL}_{sb})$ in the two figures. In the background, the contour lines of the phase $\phi_{B_s}$ are shown. The darker regions correspond to the 90\% CL range presently favoured by the experiment~\cite{hfag}. The axis of the two figures are chosen in such a way that the contour lines are the same for the degenerate and hierarchical cases.}
\label{fig:bs}
\end{figure}

In the case of $b\leftrightarrow s$ transitions, the constraints we have considered are the mass difference $\Delta m_{B_s}$ and the $\bsg$ branching ratio. We have used ref.~\cite{misiak} to compute the SM contribution to $\BRbsg$. We have constructed two separate likelihoods because  of the different $\tan\beta$ dependence of the two constraints. In fact, the $\Delta m_{B_s}$ constraint is $\tan\beta$ independent, while the $\bsg$ constraint has a linear dependence on $\tan\beta$ for moderately large $\tan\beta$.\footnote{The reason is that the leading contribution to $\BRbsg$ comes from the product of an LL insertion times an LR transition between sbottom states, which grows linearly with $\tan\beta$. At large $\tan\beta$, this dominates over the amplitude where the chiral transition occurs in the bottom quark line.} The 95\% CL contours corresponding to the two constraints are shown in \Fig{bs} for $\tan\beta = 10$. As mentioned, the $\bsg$ constraint is relevant for the LL insertions, whose contribution interferes with the SM one, but not for the RR insertions. The bounds on $\re(\delta)$ and $\im(\delta)$ in Table~\ref{tab:bounds} are obtained as in the case of $b\leftrightarrow d$ transitions. Because of the ``holes'' in the two-dimensional likelihood function shown in \Fig{bs}, the one-dimensional likelihood for $\im(\delta)$ corresponding to $\re(\delta) = 0$ has three almost disconnected parts. We calculated the bounds in Table~\ref{tab:bounds} by using the central part of the likelihood only. A comment on this procedure is in order. It is of course possible to obtain the one-dimensional likelihood for $\im(\delta)$ by a proper projection of the two-dimensional one. However, this would not take into account the fact that in the region at largest $|\im(\delta)|$ the agreement of the SM with data, $\Delta m_{B_s} \sim 2|A^\text{SM}_s|$, is reproduced through an accidental cancellation: $\Delta m_{B_s} = 2|A^\text{SM}_s + A^\text{NP}_s e^{2i\phi^\text{NP}_s}|$, where $A^\text{NP}_s e^{2i\phi^\text{NP}_s} \sim -2 A^\text{SM}_s$. Our recipe ``empirically'' discards such possibilities, and it seems appropriate for the purpose of calculating the bounds in Table~\ref{tab:bounds}. 

\begin{figure}
\begin{center}
\includegraphics[width=\textwidth]{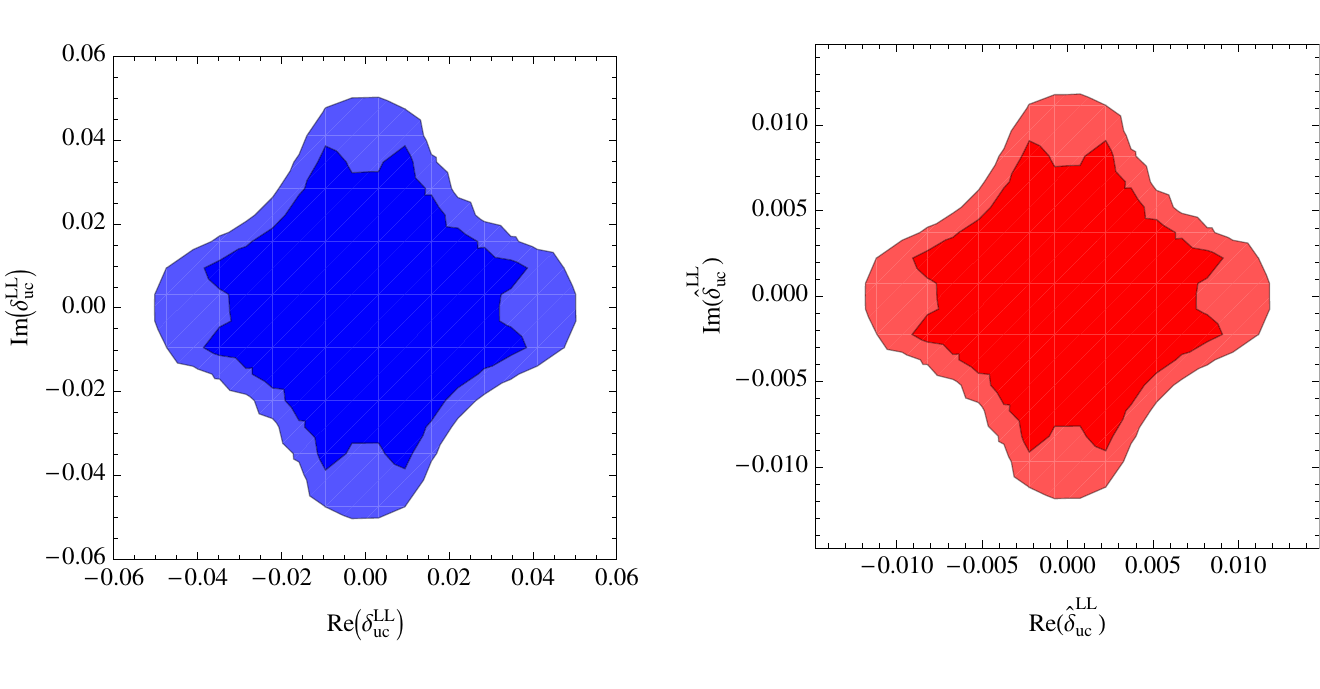}
\end{center}
\caption{95\% CL (light shading) and 68\% CL (dark shading) bounds on the real and imaginary parts of $\delta^{LL}_{uc}$ (left, blue) and $\hat\delta^{LL}_{uc}\equiv \hat\delta^{LL}_{ut}\hat\delta^{LL*}_{ct}$ (right, red) from $D^0$--$\bar D^0$ oscillations for $\tilde m = M_3  = \mu = 350\GeV$.}
\label{fig:cu}
\end{figure}

Finally, we show in \Fig{cu} the bound on the $c\leftrightarrow u$ transitions obtained from $D^0$--$\bar D^0$ mixing. The theoretical prediction for the SM contribution to the mixing amplitude is affected by a large uncertainty due to long-distance contributions and it is assumed to lie in the interval $(-0.02,0.02)\,\text{ps}^{-1}$~\cite{Ciuchini:2007cw}, with flat probability distribution. We translate in this case the likelihood in a bound on $|\delta|$ by considering the one-dimensional section of the two-dimensional likelihood along the $|\re(\delta)| = |\im(\delta)|$ line. 

\begin{table}
\begin{equation*}
\begin{array}{c|ccc|c}
\hline
\hline
\textrm{Parameter} & \textrm{Value} & \textrm{Gaussian ($\sigma$)} & \textrm{Uniform ($\frac{\Delta}{2}$)} & \textrm{Reference}\\
\hline
\hline
|\eps_K|  & 2.229  \times 10^{-3} & 0.012 \times 10^{-3} & - &  \cite{Amsler:2008zz} \\
\Delta m _K  \,  ( \textrm{ps}^{-1})  & 5.292 \times 10^{-3}  & 0.009 \times 10^{-3}  & - & \cite{Amsler:2008zz}  \\
\text{BR}(B \to X_s\gamma) & 3.55 \times 10^{-4}  & 0.26  \times 10^{-4} & - &  \cite{Barberio:2007cr} \\
\Delta m_{B_s} \, (\textrm{ps}^{-1}) & 17.77 & 0.12 &  - & \cite{Amsler:2008zz} \\
\Delta m_{B_d} \, (\textrm{ps}^{-1}) & 0.507 & 0.005 & - & \cite{Amsler:2008zz} \\
\phi_{B_d} [^\text{o}] & -4.1 & 2.1 & - & \cite{magici} \\
\left| M^D_{12} \right|  \,  ( \textrm{ps}^{-1})  & 7.7  \times 10^{-3}  & 2.5  \times 10^{-3} & - & \cite{Ciuchini:2007cw} \\
\hline
\bar{\rho} & 0.167 & 0.051 & - & \cite{Bona:2006sa} \\
\bar{\eta} & 0.386 & 0.035 & - & \cite{Bona:2006sa} \\
\lambda & 0.2255 & 0.010 & - & \cite{Amsler:2008zz}\\
|V_{cb}| & 41.2   \times 10^{-3} & 1.1  \times 10^{-3} & - & \cite{Amsler:2008zz}\\
\hline
F_K \, (\textrm{GeV})& 0.160 & - & - &  \cite{Amsler:2008zz} \\
F_{B_d}  \, (\textrm{MeV}) & 189 & 27 & -& \cite{Hashimoto:2004hn} \\
F_{B_s} \sqrt{B_s}  \, (\textrm{MeV}) & 262 & 35 & -&  \cite{Hashimoto:2004hn} \\
F_{D} \, (\textrm{MeV}) & 201 & 3& 17& \cite{Bona:2007vi}  \\
\hat{B}_K & 0.79 & 0.04 & 0.08 & \cite{Bona:2007vi}\\
B_1^B & 0.88 & 0.04 & 0.10 & \cite{Bona:2007vi} \\
\eta_{cc} & 0.47 & 0.04 & - & \cite{Buras:2008nn} \\
\eta_{ct} & 0.5765 & 0.0065 & - & \cite{Buras:2008nn} \\
\eta_{tt} & 1.43 & 0.23 & - & \cite{Buras:2008nn} \\
\hline
\end{array}
\end{equation*}
\caption{Main inputs used in the numerical analysis.}
\label{tab:inputs}
\end{table}

In the hierarchical case, the bound from the $s\leftrightarrow d$ transitions apply to the product $\hat\delta^{LL}_{db} \hat\delta^{LL*}_{sb} \equiv \hat\delta^{LL}_{ds}$. It is therefore possible to compare that bound with the indirect one obtained from the constraints on $\hat\delta^{LL}_{sb}$ and $\hat\delta^{LL}_{db}$. It turns out that the combined bound is stronger than the direct one in the case of $\Delta m_K$ but not in the case of $\epsilon_K$. 

If the parameters $\hat \delta$ are related to the hierarchy according to the relation $\hat \delta \sim {\tilde m}^2_\ell / {\tilde m}^2_h$, from the results in Table~\ref{tab:bounds} we obtain a lower bound on the heavy mass scale
\beq
{\tilde m}_h \gsim \left( \frac{{\tilde m}_\ell}{350 \ \textrm{GeV}}\right)^{1/2}~5 \ \textrm{TeV}.
\eeq
As discussed in the Appendix, it is plausible to expect that, independently of the value of the hierarchy ${\tilde m}_\ell / {\tilde m}_h$, the size of the parameters $\hat\delta^{LL}_{sb}$, $\hat\delta^{LL}_{db}$ cannot be smaller than the corresponding  CKM angles, $|V_{td}|$, $|V_{ts}|$ respectively. 
Thus, it is particularly interesting to probe experimentally flavor processes up to the level of $|\hat\delta^{LL}_{db}|\approx 8\times 10^{-3}$, $|\hat\delta^{LL}_{sb}|\approx 4\times 10^{-2}$ and $|\hat\delta^{LL}_{ds}|=|\hat\delta^{LL}_{db} \hat\delta^{LL*}_{sb} |\approx 3\times 10^{-4}$. The present constraints on the $b \leftrightarrow d$ transitions and on $\epsilon_K$ are at the edge of probing this region. An interesting conclusion is that hierarchical soft terms predict that new-physics effects in $b \leftrightarrow s$ transitions can be expected just beyond the present experimental sensitivity. 

\section{The Phase of the $B_s$ Mixing}
\label{sec:phase}

Let us now discuss the implications for the phase of the $B_s$ mixing. In the hierarchical scenario, the new-physics effects in $b\leftrightarrow s$ transitions are particularly promising. We have already pointed out that the value of $\hat\delta^{LL}_{bs}$ might be not so far from saturating the bound in Table~\ref{tab:bounds}. On top of that, a value of the insertion parameter close to its $\Delta B = 1$ bound gives rise to effects in $\Delta B = 2$ observables that are more pronounced in the hierarchical than in the degenerate case. The reason goes back to \eq{2vs1}. For most values of $\tan\beta$, the bound on the insertions is mainly due to the $\bsg$ constraint. Its translation into a constraint on $\Delta B=2$ observables such as $\Delta m_{B_s}$ or the phase $\phi_{B_s}$ of the $B_s$--$\bar B_s$ mixing depends on the scenario we consider. \Eq{2vs1} shows that for $(g^{(3)}/g^{(1)})(f/f^{(1)})^2\sim 1$ the bound on $\Delta B=2$ observables is expected to be looser in the hierarchical case. This is confirmed by the relative size of the $\Delta B=1$ and $\Delta B=2$ constraints in \Fig{bs}. 

\begin{figure}
\begin{center}
\includegraphics[width=0.66\textwidth]{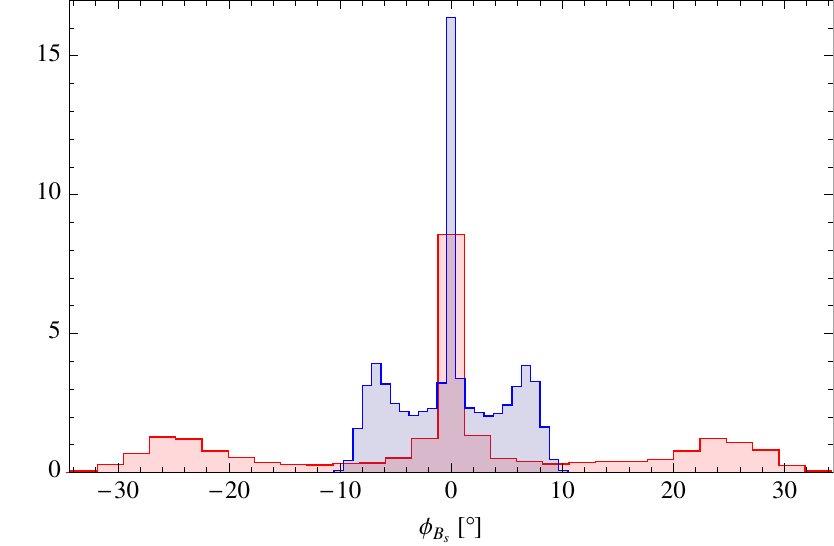}
\end{center}
\caption{Expected distribution of the phase $\phi_{Bs}$, as determined by the $\BRbsg$ and $\Delta m_{B_s}$ constraints in the degenerate (blue) and hierarchical (red), for $\tan\beta = 10$.}
\label{fig:phase}
\end{figure}

The previous considerations have interesting implications on the possible size of new-physics effects in the phase of $B_s$ mixing. The $B_s$--$\bar B_s$ mixing amplitude in the presence of new physics can be parameterized as 
\begin{equation}
\label{eq:HfullSM}
\langle B_s | H^\text{full}_\text{eff} | \bar B_s \rangle = C_{B_s} e^{2i\phi_{B_s}}
\langle B_s | H^\text{SM}_\text{eff} | \bar B_s \rangle ,
\end{equation}
where $H^\text{full}_\text{eff} = H^\text{SM}_\text{eff}+ H^\text{NP}_\text{eff}$, $\langle B_s | H^\text{SM}_\text{eff} | \bar B_s \rangle = A^\text{SM}_s e^{-2i\beta_s}$, $\langle B_s | H^\text{NP}_\text{eff} | \bar B_s \rangle = A^\text{NP}_s e^{2i(\phi^\text{NP}_s-\beta_s)}$, and $\beta_s = \arg(-(V_{ts}V^*_{tb})/(V_{cs}V_{cb}^*)) = 0.018\pm 0.001$. Recent measurements from the CDF~\cite{Aaltonen:2007he} and D0~\cite{:2008fj} collaborations have shown a mild tension between the experimental value $\phi_{B_s} \sim -20^\text{o}$ (for the allowed region closer to the origin) and its SM prediction, $\phi_{B_s} = 0^\text{o}$, at the $2.5\,\sigma$ level~\cite{dev,CKMfitter,hfag}. In the supersymmetric scenarios under consideration, the value of the phase $\phi_{B_s}$ can be read from the contour lines in  \Fig{bs}. The lines have been obtained by fixing all the relevant parameters to their central values. They converge in the two points corresponding to a vanishing total amplitude $A^\text{SM}_s + A^\text{NP}_s e^{2i\phi^\text{NP}_s}$. The figure shows that in the region allowed by both the $\BRbsg$ and $\Delta m_{B_s}$ constraints, the phase reaches larger values in the hierarchical case. This is apparent in \Fig{phase}, where the expectation for $\phi_{B_s}$  in the two scenarios has been shown in the form of an histogram (for a fixed value of $\tan\beta = 10$). The hierarchical case allows values of the phase $\phi_{B_s}$ about three times larger than in the degenerate case, in agreement with the generic expectation from \eq{2vs1}. The range of $\phi_{B_s}$ presently favored by the experiment is shown in \Fig{bs}. 

\section{Conclusions}

Hierarchical soft terms describe a class of supersymmetric theories which is characterized by the existence of two separated mass scales: a large mass ${\tilde m}_h$ for the first two generations of squarks and sleptons and a smaller mass ${\tilde m}_\ell$, of electroweak-scale size, for the rest of the spectrum. A certain hierarchy of the ratio $ {\tilde m}_h/{\tilde m}_\ell$ is not incompatible with naturalness, and it is welcome to relax constraints from $K^0$-${\bar K}^0$ mixing and $\epsilon_K$. 

This class of theories includes radical proposals in which ${\tilde m}_h$ is in the range of hundreds of TeV, fully addressing the supersymmetric flavor problem at the price of a certain amount of unnaturalness. However, the pattern of hierarchical soft terms is also useful to describe less extreme scenarios in which there is a more modest mass separation in the squark sector, nevertheless sufficient to make the degeneracy assumption a poor starting point.

Hierarchical soft terms make well-defined and interesting predictions in flavor physics. Flavor-violating effects in the down sector are described by four complex numbers: ${\hat \delta}^{LL}_{db}$, ${\hat \delta}^{LL}_{sb}$, ${\hat \delta}^{RR}_{db}$, ${\hat \delta}^{RR}_{sb}$. There are fewer free parameters than in the ordinary case of degenerate squarks, mostly because the $d \leftrightarrow s$ transition is determined by the product of $d \leftrightarrow b$ and $b \leftrightarrow s$ transitions. Also, under certain assumptions, flavor and chiral violating transitions are specified in terms of $\hat \delta$ and of the same parameters that describe squark mixing in the third generation. Another interesting peculiarity is the correlation between $\Delta F=1$ and $\Delta F=2$ transitions, which is characteristic of the hierarchical soft term pattern and distinct from the one derived in the case of degeneracy.

In this paper we have analyzed how present experiments constrain the parameters ${\hat \delta}$. The limits are derived by calculating the likelihood function for new-physics effects and combining the different experimental data and theory parameters with their relative errors. We have also applied the same procedure to the case of degeneracy, revisiting the limits on the mass insertion parameters $\delta$.

For a degenerate spectrum, the mass insertions $\delta$ are the appropriate way to parametrize new flavor-violating effects. The coefficients $\delta$ describe the small deviations from universality but, lacking the knowledge of a complete theory of soft terms, they can only be treated as free parameters and do not provide information on the required experimental sensitivity to discover new-physics effects. The analogous quantities in the hierarchical scheme, ${\hat \delta}$, are related either to the $ {\tilde m}_\ell/{\tilde m}_h$ hierarchy or to CKM angles, because of the special assumptions made on the pattern of soft terms. Therefore the quantities ${\hat \delta}$ are associated to physical parameters and they provide a defined target for the required experimental sensitivity. In particular, we expect that each ${\hat \delta}_{i3}$ is larger than the maximum between $ {\tilde m}^2_\ell/{\tilde m}^2_h$ and the CKM elements $V^*_{3i}$. The results obtained in Table~\ref{tab:bounds} show that present experiments have not yet probed $u \leftrightarrow c$ transitions at the level required by ${\hat \delta}_{i3}=V^*_{3i}$, and have only marginally tested the case of $d \leftrightarrow s$ and $d \leftrightarrow b$ transitions. 
On the other hand, experiments have begun to explore the crucial range of  values for ${\hat \delta}_{sb}$ in $s \leftrightarrow b$ transitions. In this respect, it is tantalizing that there are claims for a deviation from the SM predictions in the phase of $B_s$ mixing, $\phi_{B_s}$~\cite{dev,CKMfitter,hfag}. Hierarchical soft terms could account for such new-physics effect, compatibly with the other constraints in the $b$-$s$ system. Actually we have proved that, because of the correlation between $\Delta F=1$ and $\Delta F=2$ transitions, hierarchical soft terms can lead to larger values of $\phi_{B_s}$ than degenerate ones, for an equal value of $\tan\beta$. Independently of the reliability of the alleged anomaly in $\phi_{B_s}$, the hypothesis of hierarchical soft terms represents an interesting benchmark to confront experimental searches in flavor physics.  

\section*{Acknowledgments}
We thank Marco Ciuchini for useful discussions. 

\section*{Appendix}

In this Appendix we compute the fermion-sfermion mixing matrix $\W$ in the limit of hierarchical soft terms. We also discuss the conditions under which the heavy-squark contribution can be neglected in the amplitude of \eq{ampl2} and the natural size of the flavor-violating parameters $\hat \delta$. 

In a general basis in which the quark mass matrix is not necessarily diagonal, $\W$ is a combination of the matrices that diagonalize the quark and squark mass matrices $M$ and $\MM$ respectively,
\begin{equation}
\label{eq:Wdef}
\W = \begin{pmatrix}
U_L & 0 \\ 
0 & U_R
\end{pmatrix}
\W^\prime , 
\quad
U_R M U_L^\dagger = {\rm diagonal} ,
\quad
\W^{\prime \dagger} \MM \W^{\prime} = {\rm diagonal}.
\end{equation}

Because the relevant amplitudes will turn out to be dominated by loops with only third-generation squark exchange, we are justified to neglect chiral-violating entries in the squark mass matrix involving first or second generation indices.
Under this assumption and working at leading order in an expansion in inverse powers of the heavy-squark mass scale, we obtain
\beq
\W^\prime =\begin{pmatrix}
{\tilde U}_L & {\hat \delta}^{LL} \cos\theta & 0 &  -{\hat \delta}^{LL} \sin\theta e^{i\phi} \\
- {\hat \delta}^{LL\dagger}{\tilde U}_L & \cos\theta & 0 & -\sin\theta e^{i\phi} \\
0 & {\hat \delta}^{RR} \sin\theta e^{-i\phi} & {\tilde U}_R & {\hat \delta}^{RR} \cos\theta \\
0 &  \sin\theta e^{-i\phi} & -{\hat \delta}^{RR\dagger}{\tilde U}_R &  \cos\theta 
\end{pmatrix},
\label{eq:wpr}
\eeq
where we have omitted the generation indices of the first two generations. 
The $2\times 2$ unitary matrices ${\tilde U}_{L,R}$ diagonalize the $2\times 2$ blocks of the heavy states in the squark mass matrix (which we call $\MM_{hL}$ and $\MM_{hR}$) according to
\beq
{\tilde U}_{L}^\dagger \MM_{hL} {\tilde U}_{L} = {\rm diagonal},~~~~
{\tilde U}_{R}^\dagger \MM_{hR} {\tilde U}_{R} = {\rm diagonal}.
\eeq
The two-component vectors ${\hat \delta}^{LL,RR}_{i3}$ $(i=1,2)$ are given by
\beq
{\hat \delta}^{LL}_{i3}\equiv -\sum_{j=1}^2 \left( {{\cal M}_{hL}^{-2}} \right)_{ij} \MM_{Lj,3},~~~~
{\hat \delta}^{RR}_{i3}\equiv -\sum_{j=1}^2 \left( {{\cal M}_{hR}^{-2}} \right)_{ij} \MM_{Rj,3}.
\label{eq:defd}
\eeq
It is easy to verify that this definition coincides with \eq{defins}, at the leading order in the expansion and neglecting quark rotation effects. Finally, $\theta$ and $\phi$ are the parameters determining the diagonalization of the light-squark sector and are defined by
\beq
\tan 2\theta \equiv \frac{2 \left| \MM_{L3,R3} \right|}{\MM_{L3,L3}-\MM_{R3,R3}}, ~~~~
e^{i\phi} \equiv  \frac{\MM_{L3,R3}}{ \left| \MM_{L3,R3} \right|}.
\eeq

The result presented in the text in \eq{fLO2} can now be easily derived by replacing \eq{wpr} into \eq{ampl2}. Moreover, we can use \eq{wpr} to compare the contributions to flavor-violating amplitudes from heavy and light squarks . For instance, the flavor transition between the first and second generations in the down-left sector, obtained from \eq{ampl2}, is given by
\beq
f\left( \frac{\MM_D}{M^2}\right)_{d_Ls_L} = \frac{{\tilde m}_h^2}{M^2} \Delta_h f^{(1)}\left( \frac{{\tilde m}_h^2}{M^2}\right) +{\hat \delta}_{13}^{LL}{\hat \delta}_{23}^{LL*}f\left( \frac{{\tilde m}_\ell^2}{M^2}\right) .
\label{eq:exx}
\eeq
Here, for simplicity, we have neglected quark rotations and we have considered near degeneracy among the heavy squark states (with a common mass ${\tilde m}_h$) and among the light squark states (with a common mass ${\tilde m}_\ell$). We have defined $\Delta_h \equiv (\MM_{hL})_{12}/{\tilde m}_h^2$ to parametrize the mass insertion in the heavy sector. Using the property that, for large $x$, $f(x)\sim 1/x$ (and therefore $f^{(1)}(x)\sim 1/x^2$), we obtain that the second term in \eq{exx} dominates over the first one when
\beq
{\hat \delta}^{LL} \gsim \Delta_h^{1/2} \frac{{\tilde m}_\ell}{{\tilde m}_h}.
\label{eq:cond}
\eeq
Analogous considerations hold for ${\hat \delta}^{RR}$.
When the condition in \eq{cond} is satisfied, we are allowed to neglect the heavy-squark contribution in the loop diagram.

To establish if the condition is satisfied we have to discuss what is the natural range of values for ${\hat \delta}^{LL}$. A lower limit on ${\hat \delta}^{LL}$ is obtained from \eq{defd} with the requirement that any chiral-conserving entry of $\MM_D$ is at least of size ${\tilde m}_\ell^2$,
\beq
{\hat \delta}^{LL} \gsim \frac{{\tilde m}^2_\ell}{{\tilde m}^2_h}.
\label{eq:natd1}
\eeq
An upper limit on ${\hat \delta}^{LL}$  is derived by observing that the light left squark receives a contribution from the heavy sector to its mass square equal to
\beq
-{\hat \delta}^{LL\dagger}\MM_{hL}{\hat \delta}^{LL}\cos^2\theta 
-{\hat \delta}^{RR\dagger}\MM_{hR}{\hat \delta}^{RR}\sin^2\theta \sim \order{ {{\hat \delta}^{LL2}} {\tilde m}^2_h}.
\eeq
Thus, barring special cancellations, the hierarchical separation between the light and heavy sectors is maintained only if
\beq
{\hat \delta}^{LL} \lsim \frac{{\tilde m}_\ell}{{\tilde m}_h}.
\label{eq:natd2}
\eeq

The natural range for ${\hat \delta}^{LL}$ (or ${\hat \delta}^{RR}$) is defined by \eq{natd1} and \eq{natd2}. In the absence of any GIM suppression in the heavy sector (\emph{i.e.}\ when $\Delta_h \approx 1$), the natural values of ${\hat \delta}^{LL}$ are nearly inconsistent with the condition in \eq{cond}. However, as discussed in the text, the constraint from $\epsilon_K$ require that $\Delta_h < 10^{-2} {\tilde m}_h/(3\TeV)$. In presence of a mechanism justifying the smallness of $\Delta_h$ (like, for instance, an approximate U(2) symmetry), the condition in \eq{cond} can be satisfied.

When the ratio ${\tilde m}_h/{\tilde m}_\ell$ becomes very large, the quark rotation angles in $U_{L,R}$ can dominate over those of $\W^\prime$ in \eq{Wdef}. In this case, \eq{cond} is automatically satisfied, and the assumption of neglecting heavy squarks in the loop diagram is perfectly justified. Assuming that the CKM matrix $V=U_L^uU_L^{d\dagger}$ is dominated by the rotation in the down sector, we obtain
\beq
 {\hat \delta}^{LL}_{db} \approx V^*_{td}, ~~~~{\hat \delta}^{LL}_{sb} \approx V^*_{ts}.
 \label{eq:uff}
 \eeq
 Thus, excluding unexpected cancellations, ${\hat \delta}^{LL}$ cannot be smaller than the maximum between ${\tilde m}^2_\ell/{\tilde m}^2_h$ and what given in \eq{uff}. 
 Although we cannot directly relate $U_R$ to CKM angles, we expect that the result in \eq{uff} will hold approximately for ${\hat \delta}^{RR}$ too if, for instance, the quark mass matrix is nearly symmetric.


\begin{thebibliography}{99}

\bibitem{mfv}
  G.~D'Ambrosio, G.~F.~Giudice, G.~Isidori and A.~Strumia,
  Nucl.\ Phys.\  B {\bf 645} (2002) 155
  [arXiv:hep-ph/0207036].
  
\bibitem{dg}
  S.~Dimopoulos and H.~Georgi,
  Nucl.\ Phys.\  B {\bf 193} (1981) 150.
  
\bibitem{nirs}
  Y.~Nir and N.~Seiberg,
  Phys.\ Lett.\  B {\bf 309} (1993) 337
  [arXiv:hep-ph/9304307].
  
\bibitem{Nir:2007ac}
  Y.~Nir,
  JHEP {\bf 0705} (2007) 102
  [arXiv:hep-ph/0703235].
  
\bibitem{modali}
  M.~Leurer, Y.~Nir and N.~Seiberg,
  Nucl.\ Phys.\  B {\bf 420} (1994) 468
  [arXiv:hep-ph/9310320].
  
\bibitem{dins}
  M.~Dine, A.~Kagan and S.~Samuel,
  Phys.\ Lett.\  B {\bf 243} (1990) 250.
    
\bibitem{dimg}
  S.~Dimopoulos and G.~F.~Giudice,
  Phys.\ Lett.\  B {\bf 357} (1995) 573
  [arXiv:hep-ph/9507282].
  
\bibitem{Pomarol:1995xc}
  A.~Pomarol and D.~Tommasini,
  Nucl.\ Phys.\  B {\bf 466}, 3 (1996)
  [arXiv:hep-ph/9507462].
  
\bibitem{ckn}
  A.~G.~Cohen, D.~B.~Kaplan and A.~E.~Nelson,
  Phys.\ Lett.\  B {\bf 388} (1996) 588
  [arXiv:hep-ph/9607394].
  
\bibitem{u2}
  M.~Dine, R.~G.~Leigh and A.~Kagan,
  Phys.\ Rev.\  D {\bf 48} (1993) 4269
  [arXiv:hep-ph/9304299];
    P.~Pouliot and N.~Seiberg,
  Phys.\ Lett.\  B {\bf 318} (1993) 169
  [arXiv:hep-ph/9308363];
  R.~Barbieri, G.~R.~Dvali and L.~J.~Hall,
  Phys.\ Lett.\  B {\bf 377} (1996) 76
  [arXiv:hep-ph/9512388];
  R.~Barbieri, L.~J.~Hall, S.~Raby and A.~Romanino,
  Nucl.\ Phys.\  B {\bf 493} (1997) 3
  [arXiv:hep-ph/9610449];
  R.~Barbieri, L.~J.~Hall and A.~Romanino,
  Phys.\ Lett.\  B {\bf 401} (1997) 47
  [arXiv:hep-ph/9702315].
  
\bibitem{dev}  
  M.~Bona {\it et al.}  [UTfit Collaboration],
  arXiv:0803.0659 [hep-ph];
    Maurizio Pierini, talk at ICHEP08. 
  
\bibitem{CKMfitter}
  O.~Deschamps,
  arXiv:0810.3139 [hep-ph].

\bibitem{hfag}
  E.~Barberio {\it et al.}  [Heavy Flavor Averaging Group],
  arXiv:0808.1297 [hep-ex];

\bibitem{Barbieri:1987fn}
  R.~Barbieri and G.~F.~Giudice,
  Nucl.\ Phys.\  B {\bf 306}, 63 (1988).
  
\bibitem{2loopp}
  N.~Arkani-Hamed and H.~Murayama,
  Phys.\ Rev.\  D {\bf 56} (1997) 6733
  [arXiv:hep-ph/9703259];
    K.~Agashe and M.~Graesser,
  Phys.\ Rev.\  D {\bf 59} (1999) 015007
  [arXiv:hep-ph/9801446].
  
\bibitem{fprun}
  J.~L.~Feng, C.~F.~Kolda and N.~Polonsky,
  Nucl.\ Phys.\  B {\bf 546} (1999) 3
  [arXiv:hep-ph/9810500];
   J.~Bagger, J.~L.~Feng and N.~Polonsky,
  Nucl.\ Phys.\  B {\bf 563} (1999) 3
  [arXiv:hep-ph/9905292];
    J.~A.~Bagger, J.~L.~Feng, N.~Polonsky and R.~J.~Zhang,
  Phys.\ Lett.\  B {\bf 473} (2000) 264
  [arXiv:hep-ph/9911255].
  
\bibitem{paz}
    J.~Hisano, K.~Kurosawa and Y.~Nomura,
  Nucl.\ Phys.\  B {\bf 584} (2000) 3
  [arXiv:hep-ph/0002286].
  
\bibitem{contino}
  R.~Contino and I.~Scimemi,
  Eur.\ Phys.\ J.\  C {\bf 10} (1999) 347
  [arXiv:hep-ph/9809437].

\bibitem{split}
  N.~Arkani-Hamed and S.~Dimopoulos,
  JHEP {\bf 0506} (2005) 073
  [arXiv:hep-th/0405159];
    G.~F.~Giudice and A.~Romanino,
  Nucl.\ Phys.\  B {\bf 699} (2004) 65
   [arXiv:hep-ph/0406088];
  N.~Arkani-Hamed, S.~Dimopoulos, G.~F.~Giudice and A.~Romanino,
  Nucl.\ Phys.\  B {\bf 709} (2005) 3
  [arXiv:hep-ph/0409232].
  
\bibitem{BRS}
  A.~J.~Buras, A.~Romanino and L.~Silvestrini,
  Nucl.\ Phys.\  B {\bf 520}, 3 (1998)
  [arXiv:hep-ph/9712398].
  
\bibitem{Cohen:1996sq}
  A.~G.~Cohen, D.~B.~Kaplan, F.~Lepeintre and A.~E.~Nelson,
  Phys.\ Rev.\ Lett.\  {\bf 78} (1997) 2300
  [arXiv:hep-ph/9610252].
  
\bibitem{BBMR}
  S.~Bertolini, F.~Borzumati, A.~Masiero and G.~Ridolfi,
  Nucl.\ Phys.\  B {\bf 353}, 591 (1991).

\bibitem{greub}
  F.~Borzumati, C.~Greub, T.~Hurth and D.~Wyler,
  Phys.\ Rev.\  D {\bf 62} (2000) 075005
  [arXiv:hep-ph/9911245].

\bibitem{magici}
  M.~Bona {\it et al.}  [UTfit Collaboration],
  JHEP {\bf 0803}, 049 (2008)
  [arXiv:0707.0636 [hep-ph]];

\bibitem{magici2}
  D.~Becirevic {\it et al.},
  Nucl.\ Phys.\  B {\bf 634}, 105 (2002)
  [arXiv:hep-ph/0112303];
  M.~Ciuchini {\it et al.},
  JHEP {\bf 9810}, 008 (1998)
  [arXiv:hep-ph/9808328].
  
\bibitem{GGMS}
  F.~Gabbiani, E.~Gabrielli, A.~Masiero and L.~Silvestrini,
  Nucl.\ Phys.\  B {\bf 477} (1996) 321
  [arXiv:hep-ph/9604387].

\bibitem{misiak}
  M.~Misiak {\it et al.},
  Phys.\ Rev.\ Lett.\  {\bf 98} (2007) 022002
  [arXiv:hep-ph/0609232];
  M.~Misiak and M.~Steinhauser,
  Nucl.\ Phys.\  B {\bf 764} (2007) 62
  [arXiv:hep-ph/0609241].
    
\bibitem{Ciuchini:2007cw}
  M.~Ciuchini, E.~Franco, D.~Guadagnoli, V.~Lubicz, M.~Pierini, V.~Porretti and L.~Silvestrini,
  Phys.\ Lett.\  B {\bf 655} (2007) 162
  [arXiv:hep-ph/0703204].
  
\bibitem{Amsler:2008zz}
  C.~Amsler {\it et al.}  [Particle Data Group],
  Phys.\ Lett.\  B {\bf 667} (2008) 1.

\bibitem{Barberio:2007cr}
  E.~Barberio {\it et al.}  [Heavy Flavor Averaging Group (HFAG)
                  Collaboration],
  arXiv:0704.3575 [hep-ex].
  
\bibitem{Bona:2006sa}
  M.~Bona {\it et al.}  [UTfit Collaboration],
  Phys.\ Rev.\ Lett.\  {\bf 97}, 151803 (2006)
  [arXiv:hep-ph/0605213], as updated at \texttt{http://www.utfit.org};
J. Charles, ``CKMFitter update, short status of New Physics in $B\bar B$ mixing'', Capri 2008 conference, Anapri, Capri island (2008). To appear in the Proceedings Supplements of Nuclear Physics B. 

\bibitem{Hashimoto:2004hn}
  S.~Hashimoto,
  Int.\ J.\ Mod.\ Phys.\  A {\bf 20} (2005) 5133
  [arXiv:hep-ph/0411126].
  
\bibitem{Bona:2007vi}
  M.~Bona {\it et al.}  [UTfit Collaboration],
  JHEP {\bf 0803} (2008) 049
  [arXiv:0707.0636 [hep-ph]].
  
\bibitem{Buras:2008nn}
  A.~J.~Buras and D.~Guadagnoli,
  Phys.\ Rev.\  D {\bf 78}, 033005 (2008)
  [arXiv:0805.3887 [hep-ph]].
  
\bibitem{Misiak:2008ss}
  M.~Misiak,
  arXiv:0808.3134 [hep-ph].
  
  \bibitem{Aaltonen:2007he}
  T.~Aaltonen {\it et al.}  [CDF Collaboration],
  Phys.\ Rev.\ Lett.\  {\bf 100} (2008) 161802
  [arXiv:0712.2397 [hep-ex]].

\bibitem{:2008fj}
    V.~M.~Abazov {\it et al.}  [D0 Collaboration],
  arXiv:0802.2255 [hep-ex].

\end{thebibliography}
\end{document}